\documentclass[12pt]{article}
\usepackage{eurosym}
\usepackage{amssymb}
\usepackage{amsmath}
\usepackage{cite}
\usepackage{graphicx}
\usepackage{epsf}
\usepackage{lscape}

\begin{document}

\title{Lie Symmetries for the Shallow Water Magnetohydrodynamics Equations
in a Rotating Reference Frame}
\author{Andronikos Paliathanasis$^{1,2,3}$\thanks{%
Email: anpaliat@phys.uoa.gr} \ and \ Amlan Halder$^{3}$ \\
{\ \textit{$^{1}$ Institute of Systems Science, Durban University of
Technology, }}\\
{\ \textit{PO Box 1334, Durban 4000, Republic of South Africa}} \\
{\ \textit{$^{2}$ Departamento de Matem\'{a}ticas, Universidad Cat\'{o}lica
del Norte, }} \\
{\ \textit{Avda. Angamos 0610, Casilla 1280 Antofagasta, Chile}} \\
{\ \textit{$^{3}$ School of Sciences, Woxsen University,}} \\
{\ \textit{Hyderabad 502345, Telangana, India }} }
\maketitle

\begin{abstract}
We perform a detailed Lie symmetry analysis for the hyperbolic system of
partial differential equations that describe the one-dimensional Shallow
Water magnetohydrodynamics equations within a rotating reference frame. We
consider a relaxing condition $\mathbf{\mathbf{\nabla }}\left( h\mathbf{B}%
\right) \neq 0$ for the one-dimensional problem, which has been used to
overcome unphysical behaviors. The hyperbolic system of partial differential
equations depends on two parameters: the constant gravitational potential $g$
and the Coriolis term $f_{0}$, related to the constant rotation of the
reference frame. For four different cases, namely $g=0,~f_{0}=0$; $g\neq
0\,,~f_{0}=0$; $g=0$, $f_{0}\neq 0$; and $g\neq 0$, $f_{0}\neq 0$ the
admitted Lie symmetries for the hyperbolic system form different Lie
algebras. Specifically the admitted Lie algebras are the $L^{10}=\left\{
A_{3,3}\rtimes A_{2,1}\right\} \otimes _{s}A_{5,34}^{a}$; $%
L^{8}=A_{2,1}\rtimes A_{6,22}$; $L^{7}=A_{3,5}\rtimes\left\{ A_{2,1}\rtimes
A_{2,1}\right\} $; and $L^{6}=A_{3,5}\rtimes A_{3,3}~$respectively, where we
use the Morozov-Mubarakzyanov-Patera classification scheme. For the general
case where $f_{0}g\neq 0$, we derive all the invariants for the Adjoint
action of the Lie algebra $L^{6}$ and its subalgebras, and we calculate all
the elements of the one-dimensional optimal system. These elements are then
considered to define similarity transformations and construct analytic
solutions for the hyperbolic system.

\bigskip

\end{abstract}

\section{Introduction}

\label{sec1}

Saint-Venant in \cite{sv} introduced a set of partial differential equations
to describe the flow of a thin layer of inviscid fluid. These equations form
a hyperbolic system known as the Shallow-Water (SW) equations. The SW
equations can model water flow in rivers, lakes, and other physical systems
where the flow creates a thin layer, allowing the vertical fluid movements
to be neglected. For various applications of the SW equations, we refer the
reader to \cite{swb1,swb2,swb3,swb4,swb5,swb6} and the references therein.
In an ideal physical system, the SW equations describe the conservation of
mass and momentum. In a real physical system, gravitational force terms are
introduced in the momentum equation, along with the Coriolis force,
friction, and other effects \cite{swb2}. On the other hand, in multi-fluid
flows with nonzero interaction and energy transfer between the two fluids,
the equation of motion for the mass is modified \cite{fe1,fe2}.

The Magnetohydrodynamic Shallow-Water system (SWMHD) was introduced to
describe the solar tachocline \cite{gib}, a thin layer in the Sun that marks
the transition in angular velocity behavior between the radiative zone and
the convection zone \cite{sol1}. The tachocline is believed to play an
important role in the evolution of solar behavior and structure. Moreover,
it has been proposed as the main mechanism for explaining the solar dynamo;
for more details, see \cite{sb1}. In \cite{sb2}, it was shown that SW waves
in the tachocline lead to a dispersion equation that can describe Alfven
waves and magneto-gravity waves \cite{sb2}. The Coriolis term in the
rotating reference frame results in waves that do not change shape, while in
the absence of the Coriolis force, singularities in the waves appear. In
\cite{sb3}, slow and fast magnetic Rossby waves were identified for the
tachocline. The SWMHD system has been applied to various physical systems
beyond the tachocline \cite{sb4,sb5,sb6}.

In this study, we focus on the algebraic properties of the SWMHD system
within a rotating reference frame and under a constant gravitational force.
Specifically, we apply Lie symmetry analysis \cite%
{ibra,Bluman,Stephani,olver} to determine all the point transformations that
leave the SWMHD system invariant. Symmetry analysis is a crucial approach in
the study of nonlinear differential equations \cite{sa1,sa2,sa3,sa4,sa5,sa6}
and has various applications in SW models \cite%
{ls1,ls2,ls4,ls4,ls5,ls5a,ls5b,ls5c} as well as MHD equations \cite%
{ls6,ls7,ls8,ls9,ls10,ls11,ls12}. The Lie symmetry analysis for the
one-dimensional SWMHD equations was recently presented in \cite{smel1}. In
\cite{smel1}, a non-rotating reference frame was assumed, and Gauss' law for
the magnetic field was imposed. However, as demonstrated, this requirement
for the one-dimensional system can lead to unphysical results, and this
constraint can be relaxed when studying the one-dimensional system \cite%
{bbr,bbr1,bbr2,bbr3}. The inclusion of this term restores Galilean
invariance for the SWMHD equations.

The structure of the paper is as follows:

In Section \ref{sec2}, we present the basic definitions and properties of
Lie symmetry analysis. In particular, we introduce the constraint equation,
where a given differential equation remains invariant under the action of an
infinitesimal one-parameter point transformation. We discuss the main
application of Lie symmetry analysis, which is the concept of similarity
transformations, and how these transformations can be applied to reduce a
differential equation. Furthermore, we outline the main steps for
calculating the one-dimensional system for the admitted Lie symmetries of
the differential equation.

In Section \ref{sec3}, we introduce the SWMHD system in the presence of a
gravitational field and in a rotating reference frame, which leads to the
appearance of a Coriolis force. Since we focus on the one-dimensional
system, we consider a set of five hyperbolic partial differential equations.
We also consider a system with a relaxed condition for Gauss' law for the
magnetic field, as this leads to physically acceptable solutions for the
one-dimensional system.

The Lie symmetry classification problem for the SWMHD equations is presented
in Section \ref{sec4}. Additionally, in Sections \ref{sec5} and \ref{sec6},
we present the derivation of the one-dimensional optimal system and its
application in constructing similarity solutions, respectively. Finally, in
Section \ref{sec7}, we provide our conclusions.

\section{Preliminaries}

\label{sec2}

In this section, we introduce the fundamental concepts and properties of Lie
symmetry analysis for differential equations. We explore the use of
similarity transformations and their application in deriving solutions.
Finally, we define the one-dimensional optimal system.

\subsection{Lie symmetries}

Consider the independent variables $y^{i}$ and the dependent variables $%
\Psi^{A}$, which are defined in the jet space $B = B(y^{i}, \Psi^{A},
\Psi_{,i}^{A},...)$. We note that in this work, we use Einstein's summation
convention.

Let the function $\mathbf{H} = \mathbf{H}(y^{i}, \Psi^{A}, \Psi_{,i}^{A})$
describe a system of first-order differential equations. In this study, we
focus on the case of first-order differential equations, but the following
analysis can also be extended to higher-dimensional differential equations.

Under the application of the one-parameter infinitesimal point transformation

\begin{align}
\bar{y}^{i}& =y^{i}+\varepsilon \xi ^{i}(y^{k},\Psi ^{B})~,  \label{pr.01} \\
\bar{\Psi}^{A}& =\Psi ^{A}+\varepsilon \eta ^{A}(y^{k},\Psi ^{B})~,
\label{pr.02}
\end{align}%
with $\varepsilon ^{2}\rightarrow 0$, the set of differential equations~$%
\mathbf{H}$, becomes%
\begin{equation}
\mathbf{\bar{H}}=\mathbf{\bar{H}}\left( \bar{y}^{i}\left( \varepsilon
,y^{k},\Psi ^{B}\right) ,\bar{\Psi}^{A}\left( \varepsilon ,y^{k},\Psi
^{B}\right) ,\bar{\Psi}_{,i}^{A}\left( \varepsilon ,y^{k},\Psi ^{B}\right)
\right) .
\end{equation}%
Thus, the set of differential equations $\mathbf{H}$ remain invariant under
the application of the infinitesimal transformation (\ref{pr.01}),~(\ref%
{pr.02}) if and only if \cite{Stephani,Bluman}%
\begin{equation}
\mathbf{\bar{H}}\left( \bar{y}^{i}\left( \varepsilon ,y^{k},\Psi ^{B}\right)
,\bar{\Psi}^{A}\left( \varepsilon ,y^{k},\Psi ^{B}\right) ,\bar{\Psi}%
_{,i}^{A}\left( \varepsilon ,y^{k},\Psi ^{B}\right) \right) =\mathbf{H}%
(y^{i},\Psi ^{A},\Psi _{,i}^{A}),
\end{equation}%
that is \cite{Stephani,Bluman}%
\begin{equation}
\lim_{\varepsilon \rightarrow 0}\frac{\mathbf{\bar{H}}\left( \bar{y}%
^{i}\left( \varepsilon ,y^{k},\Psi ^{B}\right) ,\bar{\Psi}^{A}\left(
\varepsilon ,y^{k},\Psi ^{B}\right) ,\bar{\Psi}_{,i}^{A}\left( \varepsilon
,y^{k},\Psi ^{B}\right) \right) -\mathbf{H}(y^{i},\Psi ^{A},\Psi _{,i}^{A})}{%
\varepsilon }=0\text{.}  \label{sm.01}
\end{equation}

When the latter condition is true we shall say that the generator%
\begin{equation}
\mathbf{X}=\frac{\partial \bar{y}^{i}}{\partial \varepsilon }\partial _{i}+%
\frac{\partial \bar{\Psi}^{A}}{\partial \varepsilon }\partial _{A},
\end{equation}
of the infinitesimal transformation (\ref{pr.01})-(\ref{pr.02}) is a Lie
symmetry vector for the system of differential equations $\mathbf{H}$.

The symmetry condition (\ref{sm.01}) can be written in the equivalent form
\cite{Stephani,Bluman}%
\begin{equation}
\mathbf{X}^{\left[ 1\right] }\left( \mathbf{H}\right) =0,
\end{equation}%
or
\begin{equation}
\mathbf{X}^{[1]}\left( \mathbf{H}\right) =\lambda \mathbf{H}~,~\mod\mathbf{H}%
=0,  \label{sm.02}
\end{equation}%
in which $\mathbf{X}^{[1]}$ is the first extension/prolongation vector of $%
\mathbf{X}$ in the jet space, and $\lambda $ is a function which should
derived.

Condition (\ref{sm.02}) is referred to as the Lie symmetry condition, which
leads to a system of linear differential equations for the coefficient
functions $\xi ^{i}(y^{k},\Psi ^{B})$ and $\eta ^{A}(y^{k},\Psi ^{B})$ of
the generator vector $\mathbf{X}$.

The prolongation vector $\mathbf{X}^{[1]}$ is given by the following
expression%
\begin{equation}
\mathbf{X}^{[1]}=\mathbf{X}+\eta ^{\left[ 1\right] A}\left( x^{k},\Psi
^{B},\Psi _{,i}^{B}\right) \partial _{u_{,i}^{A}},
\end{equation}%
with%
\begin{equation}
\eta ^{\left[ 1\right] A}\left( x^{k},\Psi ^{B},\Psi _{,i}^{B}\right)
=D_{i}\left( \eta ^{A}\right) -\Psi _{,j}^{A}D_{i}\xi ^{j},  \label{sm.03}
\end{equation}%
in which $D_{i}$ is the total derivative operator, that is,%
\begin{equation}
D_{i}=\frac{\partial }{\partial x^{i}}+\Psi _{,i}^{B}\frac{\partial }{%
\partial \Psi ^{B}}+...~.
\end{equation}

Therefore, $\eta ^{\left[ 1\right] }$ is expressed as%
\begin{equation}
\eta _{\left[ i\right] }^{A}=\eta _{,i}^{A}+\Psi _{,i}^{B}\eta _{,B}^{A}-\xi
_{,i}^{j}\Psi _{,j}^{A}-\Psi _{,i}^{A}\Psi _{,j}^{B}\xi _{,B}^{j}~.
\end{equation}

\subsection{Similarity transformation}

Assume now that $X=\xi ^{i}\left( y^{k},\eta ^{B}\right) \partial _{i}$\ is
a Lie symmetry vector for the system of differential equations~$%
H=H(y^{i},\Psi ^{A},\Psi _{,i}^{A})$. Then, there exist a set of variables $%
y^{i}\rightarrow z^{\beta }$, so that the Lie symmetry vector to read \cite%
{Stephani,Bluman}%
\begin{equation}
\mathbf{X}=\partial _{\beta },
\end{equation}%
which are known as normal coordinates. Hence the set of differential
equations in the new variables are defined as
\begin{equation}
\mathbf{H}=\mathbf{H}(z^{\beta },\Psi ^{A},\Psi _{,\beta }^{A}).
\end{equation}%
This change of variables is known as a similarity transformation, which, in
the case of partial differential equations, is applied to reduce the number
of independent variables.

In essence, the requirement for a similarity transformation implies that the
solution of the differential equations remains invariant along the surface
defined by the Lie symmetry vector.

If $\Psi ^{A}\left( y^{k}\right) $\ is a solution for the differential
equation, then $D_{i}\Psi ^{A},$\ is also a solution. Moreover, if $X=\xi
^{i}\partial _{i}+\eta ^{A}\partial _{A}$\ is a Lie symmetry for the set of
differential equations $H$, then by definition if $\Psi ^{A}\left(
y^{k}\right) $\ is a solution, the $X\left( \Psi ^{A}\left( y^{k}\right)
\right) $\ remain a solution, i.e. $X\left( \Psi ^{A}\left( y^{k}\right)
\right) =\bar{\Psi}^{A}\left( y^{k}\right) $, where $\bar{\Psi}^{A}\left(
y^{k}\right) \,$is also a solution.

Therefore, the operator $X^{\prime }=X-f^{i}D_{i}$\ is also a symmetry
vector, where $f^{k}$\ is an arbitrary function. Thus, for $f^{k}=\xi ^{k}$\
the vector field $X^{\prime }$\ is simplified as follows \cite%
{Stephani,Bluman}
\begin{equation}
\mathbf{X}^{\prime }\mathbf{=}\left( \eta ^{A}-\xi ^{k}\Psi _{,k}\right)
\partial _{A}.
\end{equation}%
Therefore, we can define the operator
\begin{equation}
\eta ^{A}-\xi ^{k}\Psi _{,k}=a_{0}\Psi .
\end{equation}%
This condition results in the derivation of the similarity transformation
associated with the vector field $X$. The two approaches are equivalent,
meaning that for a given symmetry vector, the corresponding similarity
transformation is unique.

However, not all symmetry vectors lead to similarity transformations that
describe independent solutions. In the following, we define the
one-dimensional optimal system, which is essential for classifying all
independent solutions that arise from the application of similarity
transformations. For the derivation of the one-optimal system for a given
algebra we refer the reader to \cite{olver}

\section{Shallow Water Magnetohydrodynamics}

\label{sec3}

The equations of SWMHD\ for a hyperbolic system of partial differential
equations, that is \cite{gib,bb0a,bb0},%
\begin{eqnarray}
h_{,t}+\nabla \left( h\mathbf{u}\right) &=&0,  \label{sw.01} \\
\left( h\mathbf{u}\right) _{,t}+\nabla \left( h\mathbf{u\mathbf{u}-}h\mathbf{%
BB+}\frac{g}{2}h^{2}\right) +2\mathbf{\Omega }\times h\mathbf{u} &=&0,
\label{sw.02} \\
\left( h\mathbf{B}\right) _{,t}+\nabla \left( h\mathbf{B\mathbf{u}-}h\mathbf{%
uB}\right) +\mathbf{\mathbf{u\nabla }}\left( h\mathbf{B}\right) &=&0,
\label{sw.03}
\end{eqnarray}%
where $h$ is the layer depth of the incompressible, perfectly conducting
fluid with an equation of state parameter $p = \frac{1}{2}gh^{2}$. The
vector field $\mathbf{u}$ represents the horizontal component of the
velocity, $\mathbf{B}$ is the magnetic field, $g$ is the gravitational
constant, and $\mathbf{\Omega}$, originating from the rotating frame,
provides the Coriolis force term.

From Maxwell's equations, we obtain the constraint $\mathbf{\nabla} \left(h
\mathbf{B}\right) = 0$. However, enforcing this constraint leads to
degenerate numerical solutions for the Riemann problem with one-dimensional
initial conditions \cite{bb1,bb2,bb3,bb3a}. To address this issue, a relaxed
condition $\mathbf{\nabla} \left(h \mathbf{B}\right) \neq 0$ has been
introduced in several studies for the one-dimensional problem. This
modification helps restore the Galilean invariance of the equations \cite%
{bb4} and can lead to physically acceptable solutions for the
one-dimensional Riemann problem \cite{bb3}. For more details, we refer the
reader to the discussions in \cite{bb5,bb6}.

Assuming that the dependent variables are functions of time $t$ and the
spatial coordinate $x$, the components of the horizontal velocity are $%
\mathbf{u} = \left( u(t,x), v(t,x) \right)^{T}$, and the components of the
magnetic field are $\mathbf{B} = \left( a(t,x), b(t,x) \right)^{T}$.

From (\ref{sw.01}), (\ref{sw.02}), and (\ref{sw.03}), we obtain the
hyperbolic system of five partial differential equations \cite{bb6}.
\begin{eqnarray}
h_{,t}+\left( hu\right) _{,x} &=&0,  \label{ww.01} \\
\left( hu\right) _{,t}+\left( hu^{2}+\frac{g}{2}h^{2}-ha^{2}\right)
_{,x}+f_{0}hv &=&0,  \label{ww.02} \\
\left( hv\right) _{,t}+\left( huv-hab\right) _{,x}-f_{0}hu &=&0,
\label{ww.03} \\
\left( ha\right) _{,t}+u\left( ha\right) _{,x} &=&0,  \label{ww.04} \\
\left( hb\right) _{,t}+\left( h\left( ub-va\right) \right) _{,x}+v\left(
ha\right) _{,x} &=&0,  \label{ww.05}
\end{eqnarray}%
where $\Omega =-\frac{1}{2}f_{0}$.

In the following we proceed with the symmetry analysis for the hyperbolic
system (\ref{ww.01}), (\ref{ww.02}), (\ref{ww.03}), (\ref{ww.04}) and (\ref%
{ww.05}).

We consider the infinitesimal one-parameter point transformation%
\begin{eqnarray}
\bar{t} &=&t+\varepsilon \xi ^{t}\left( t,x,h,\mathbf{u,B}\right) ,~\bar{x}%
=x+\varepsilon \xi ^{x}\left( t,x,h,\mathbf{u,B}\right) , \\
\bar{u} &=&u+\varepsilon \eta ^{u}\left( t,x,h,\mathbf{u,B}\right) ~,~\bar{v}%
=v+\varepsilon \eta ^{v}\left( t,x,h,\mathbf{u,B}\right) , \\
\bar{a} &=&a+\varepsilon \eta ^{a}\left( t,x,h,\mathbf{u,B}\right) ,~\bar{b}%
=b+\varepsilon \eta ^{b}\left( t,x,h,\mathbf{u,B}\right) , \\
\bar{h} &=&h+\varepsilon \eta ^{h}\left( t,x,h,\mathbf{u,B}\right) ,
\end{eqnarray}%
with generator the vector field
\begin{equation}
X=\xi ^{t}\partial _{t}+\xi ^{x}\partial _{x}+\eta ^{h}\partial _{h}+\eta
^{u}\partial _{u}+\eta ^{v}\partial _{v}+\eta ^{a}\partial _{a}+\eta
^{b}\partial _{b}.
\end{equation}

The Lie symmetry condition results in a set of constraint equations for the
coefficients of the vector field $\mathbf{X}$. The constraint system and the
admitted Lie symmetries depend on the presence of the gravitational field
and the Coriolis force. We examine all possible cases separately.

\section{Lie symmetry analysis for the SWMHD\ system}

\label{sec4}

In the following lines we present the Lie symmetry conditions (\ref{sm.02})\
for the SWMHD\ system. The admitted Lie symmetries for the SWMHD system
depend on the value of the free parameters $f_{0}$\ and $g$. It follows that
there are four different cases for the values of the gravitational field and
of the Coriolis term. Specifically, for the following four cases
(a)\thinspace $g=0,~f_{0}=0$, (b)$g\neq 0\,,~f_{0}=0$, (c) $g=0$, $f_{0}\neq
0$\ and (d) $g\neq 0$, $f_{0}\neq 0$, it follows that the admitted Lie
symmetries for the SWMHD system (\ref{ww.01}), (\ref{ww.02}), (\ref{ww.03}),
(\ref{ww.04}) and (\ref{ww.05}) form different dimensional Lie algebra. The
fist two cases, describe nonrotating systems with or without gravitational
field, while cases (c) and (d) describe rotating systems with or without
gravitational field.

\subsection{Free system}

Let us now assume that the SWMHD equations are free, that is, there is not a
gravitational field, i.e. $g=0$, and there is not any Coriolis force term,
i.e. $f_{0}=0$. Therefore, the hyperbolic system describes the SWMHD
equations is simplified as follows%
\begin{eqnarray}
h_{,t}+\left( hu\right) _{,x} &=&0, \\
\left( hu\right) _{,t}+\left( hu^{2}-ha^{2}\right) _{,x} &=&0, \\
\left( hv\right) _{,t}+\left( huv-hab\right) _{,x} &=&0, \\
\left( ha\right) _{,t}+u\left( ha\right) _{,x} &=&0, \\
\left( hb\right) _{,t}+\left( h\left( ub-va\right) \right) _{,x}+v\left(
ha\right) _{,x} &=&0,
\end{eqnarray}

The Lie symmetry conditions for this system are%
\begin{equation*}
\xi ^{t}=\xi ^{t}\left( t\right) ,~\xi ^{x}=\xi ^{x}\left( t,x\right) ,~\eta
^{h}=\eta ^{h}\left( h\right) ,
\end{equation*}%
\begin{equation}
\eta ^{u}=\left( \xi _{,x}^{x}-\xi _{,t}^{t}\right) u+\xi _{,t}^{x}~,~\eta
^{v}=\eta ^{v}\left( u,v\right) ,
\end{equation}%
\begin{equation}
\eta ^{a}=-a\left( \xi _{,t}^{t}-\xi _{,x}^{x}\right) ,~\eta ^{b}=\eta
^{b}\left( h,a,b\right) ,
\end{equation}%
\begin{equation}
\xi _{,tt}^{t}=0,~\xi _{,tt}^{x}=0,~\xi _{,xx}^{x}=0,~\xi _{,tx}^{x}=0,
\end{equation}%
\begin{equation}
\eta _{,uu}^{v}=0,~\eta _{,vv}^{v}=0,~\eta _{,uv}^{v}=0,
\end{equation}%
\begin{equation}
h\eta _{,h}^{h}-\eta ^{h}=0~,a\eta _{,a}^{b}-\left( 2a\eta _{,u}^{v}-\eta
^{b}+\eta _{,v}^{v}b\right) =0,
\end{equation}%
\begin{equation}
\eta _{,b}^{b}-\eta _{,v}^{v}=0,~h\eta _{,h}^{b}-\left( a\eta _{,u}^{v}-\eta
^{b}+\eta _{,v}^{v}b\right) =0.
\end{equation}

The solution of this system provides the generic generator $X$ for the
infinitesimal transformations that leave the SWMHD equations invariant. The
arbitrary constants in the generic vector field $\mathbf{X}$ determine the
specific admitted Lie symmetries for the hyperbolic system.

\subsubsection{Lie symmetries}

The solution of the symmetry conditions for the free SWMHD hyperbolic
equations provide a ten-dimensional Lie algebra $L^{10}~$consisted by the
vector fields%
\begin{eqnarray*}
X_{1} &=&\partial _{t}~,~X_{2}=\partial _{x}~,~X_{3}=t\partial
_{t}+x\partial _{x},~X_{4}=h\partial _{h}~,~X_{5}=t\partial _{x}+\partial
_{u},~X_{6}=\partial _{v}~, \\
X_{7} &=&u\partial _{v}+a\partial _{b}~,~X_{8}=v\partial _{v}+b\partial
_{b}~,X_{9}=t\partial _{t}-\left( u\partial _{u}+a\partial _{a}\right)
~,~X_{10}=\frac{1}{ah}\partial _{b}.
\end{eqnarray*}%
where we see that the Galileon symmetries are preserved to the system.

For the admitted vector fields we can define the corresponding one-parameter
point transformations, indeed we derive%
\begin{eqnarray*}
X_{1} &:&\bar{t}=t+\varepsilon _{1}, \\
X_{2} &:&\bar{x}=x+\varepsilon _{2}, \\
X_{3} &:&\bar{t}=e^{\varepsilon _{3}}t~,~\bar{x}=e^{\varepsilon _{3}}x, \\
X_{4} &:&\bar{h}=e^{\varepsilon _{4}}h, \\
X_{5} &:&\bar{t}=e^{\varepsilon _{5}}t,~\bar{u}=e^{\varepsilon _{5}}u, \\
X_{6} &:&\bar{v}=v+\varepsilon \\
X_{7} &:&\bar{v}=v+\varepsilon _{7}u,~\bar{b}=b+\varepsilon _{7}a, \\
X_{8} &:&\bar{v}=e^{\varepsilon _{8}}v,~\bar{b}=e^{\varepsilon _{8}}b, \\
X_{9} &:&\bar{t}=e^{\varepsilon _{9}}t,~\bar{u}=e^{-\varepsilon _{9}}u,~\bar{%
a}=e^{-\varepsilon _{9}}a \\
X_{10} &:&\bar{b}=b+\varepsilon _{10}ah\text{.}
\end{eqnarray*}

\textbf{In order to understand the algebraic properties of the admitted
symmetries we derive the commutators and the adjoint representation }$%
Ad\left( e^{\left( \varepsilon \mathbf{X}_{i}\right) }\right) $\textbf{. W
remark that the Adjoint operator is defined by the Lie bracket }$\left[ ,%
\right] $\textbf{\ as follows }%
\begin{equation}
Ad\left( \exp \left( \epsilon X_{A}\right) \right) X_{B}=X_{B}-\epsilon
\left[ X_{A},X_{B}\right] +\frac{1}{2}\epsilon ^{2}\left[ X_{A},\left[
X_{A},X_{B}\right] \right] +...~.  \label{sw.07}
\end{equation}%
\textbf{The one-dimensional subalgebras of the admitted Lie symmetries which
are not related through the adjoint representation form the one-dimensional
optimal system. The determination of the one-dimensional system is essential
in order to perform a complete classification of all the possible similarity
transformations and solutions.}$~$

The commutators for the ten-dimensional Lie algebra $L^{10}$ are presented
in Table \ref{tabl1}, while the adjoint presentation table is given in
Tables \ref{tab2} and \ref{tab2b}. Therefore from the commutators we infer
that the admitted $L^{10}~$Lie Algebra is the $\left\{ A_{3,3}\rtimes
A_{2,1}\right\} \rtimes A_{5,34}^{a}\,$\ in the Morozov-Mubarakzyanov-Patera
classification scheme \cite{m1,m2,m3,m4,m5}.

\begin{table}[tbp] \centering%
\caption{Commutators of the ten admitted Lie symmetries for the free SWMHD
hyperbolic system.}%
\begin{tabular}{ccccccccccc}
\hline\hline
$\left[ \mathbf{X}_{i},\mathbf{X}_{j}~\right] $ & $\mathbf{X}_{1}$ & $%
\mathbf{X}_{2}$ & $\mathbf{X}_{3}$ & $\mathbf{X}_{4}$ & $\mathbf{X}_{5}$ & $%
\mathbf{X}_{6}$ & $\mathbf{X}_{7}$ & $\mathbf{X}_{8}$ & $\mathbf{X}_{9}$ & $%
\mathbf{X}_{10}$ \\
$\mathbf{X}_{1}$ & $0$ & $0$ & $X_{1}$ & $0$ & $X_{2}$ & $0$ & $0$ & $0$ & $%
X_{1}$ & $0$ \\
$\mathbf{X}_{2}$ & $0$ & $0$ & $X_{2}$ & $0$ & $0$ & $0$ & $0$ & $0$ & $0$ &
$0$ \\
$\mathbf{X}_{3}$ & $-X_{1}$ & $-X_{2}$ & $0$ & $0$ & $0$ & $0$ & $0$ & $0$ &
$0$ & \thinspace $0$ \\
$\mathbf{X}_{4}$ & $0$ & $0$ & $0$ & $0$ & $0$ & $0$ & $0$ & $0$ & $0$ & $%
-X_{10}$ \\
$\mathbf{X}_{5}$ & $-X_{2}$ & $0$ & $0$ & $0$ & $0$ & $0$ & $X_{6}$ & $0$ & $%
-X_{5}$ & $0$ \\
$\mathbf{X}_{6}$ & $0$ & $0$ & $0$ & $0$ & $0$ & $0$ & $0$ & $X^{6}$ & $0$ &
$0$ \\
$\mathbf{X}_{7}$ & $0$ & $0$ & $0$ & $0$ & $-X_{6}$ & $0$ & $0$ & $X^{7}$ & $%
X^{7}$ & $0$ \\
$\mathbf{X}_{8}$ & $0$ & $0$ & $0$ & $0$ & $0$ & $-X^{6}$ & $-X^{7}$ & $0$ &
$0$ & $-X_{10}$ \\
$\mathbf{X}_{9}$ & $-X_{1}$ & $0$ & $0$ & $0$ & $X_{5}$ & $0$ & $-X^{7}$ & $%
0 $ & $0$ & $X_{10}$ \\
$\mathbf{X}_{10}$ & $0$ & $0$ & $0$ & $X_{10}$ & $0$ & $0$ & $0$ & $X_{10}$
& $-X_{10}$ & $0$ \\ \hline\hline
\end{tabular}%
\label{tabl1}%
\end{table}%

\begin{table}[tbp] \centering%
\caption{Adjoint representation of the ten admitted Lie symmetries for the
free SWMHD hyperbolic system (1/2).}%
\begin{tabular}{cccccc}
\hline\hline
$Ad\left( e^{\left( \varepsilon \mathbf{X}_{i}\right) }\right) \mathbf{X}%
_{j} $ & $\mathbf{X}_{1}$ & $\mathbf{X}_{2}$ & $\mathbf{X}_{3}$ & $\mathbf{X}%
_{4}$ & $\mathbf{X}_{5}$ \\
$\mathbf{X}_{1}$ & $X_{1}$ & $X_{2}$ & $X_{3}-\varepsilon X_{1}$ & $X_{4}$ &
$X_{5}-2X_{2}$ \\
$\mathbf{X}_{2}$ & $X_{1}$ & $X_{2}$ & $X_{3}-\varepsilon X_{2}$ & $X_{4}$ &
$X_{5}$ \\
$\mathbf{X}_{3}$ & $e^{\varepsilon }X_{1}$ & $e^{\varepsilon }X_{2}$ & $%
X_{3} $ & $X_{4}$ & $X_{5}$ \\
$\mathbf{X}_{4}$ & $X_{1}$ & $X_{2}$ & $X_{3}$ & $X_{4}$ & $X_{5}$ \\
$\mathbf{X}_{5}$ & $X_{1}+\varepsilon X_{2}$ & $X_{2}$ & $X_{3}$ & $X_{4}$ &
$X_{5}$ \\
$\mathbf{X}_{6}$ & $X_{1}$ & $X_{2}$ & $X_{3}$ & $X_{4}$ & $X_{5}$ \\
$\mathbf{X}_{7}$ & $X_{1}$ & $X_{2}$ & $X_{3}$ & $X_{4}$ & $%
X_{5}+\varepsilon X^{6}$ \\
$\mathbf{X}_{8}$ & $X_{1}$ & $X_{2}$ & $X_{3}$ & $X_{4}$ & $X_{5}$ \\
$\mathbf{X}_{9}$ & $e^{\varepsilon }X_{1}$ & $X_{2}$ & $X_{3}$ & $X_{4}$ & $%
e^{-\varepsilon }X_{5}$ \\
$\mathbf{X}_{10}$ & $X_{1}$ & $X_{2}$ & $X_{3}$ & $X_{4}-\varepsilon X_{10}$
& $X_{5}$ \\ \hline\hline
\end{tabular}%
\label{tab2}%
\end{table}%

\begin{table}[tbp] \centering%
\caption{Adjoint representation of the ten admitted Lie symmetries for the
free SWMHD hyperbolic system (2/2).}%
\begin{tabular}{cccccc}
\hline\hline
$Ad\left( e^{\left( \varepsilon \mathbf{X}_{i}\right) }\right) \mathbf{X}%
_{j} $ & $\mathbf{X}_{6}$ & $\mathbf{X}_{7}$ & $\mathbf{X}_{8}$ & $\mathbf{X}%
_{9}$ & $\mathbf{X}_{10}$ \\
$\mathbf{X}_{1}$ & $X_{6}$ & $X_{7}$ & $X_{8}$ & $X_{9}-\varepsilon X_{1}$ &
$X_{10}$ \\
$\mathbf{X}_{2}$ & $X_{6}$ & $X_{7}$ & $X_{8}$ & $X_{9}$ & $X_{10}$ \\
$\mathbf{X}_{3}$ & $X_{6}$ & $X_{7}$ & $X_{8}$ & $X_{9}$ & $X_{10}$ \\
$\mathbf{X}_{4}$ & $X_{6}$ & $X_{7}$ & $X_{8}$ & $X_{9}$ & $e^{\varepsilon
}X_{10}$ \\
$\mathbf{X}_{5}$ & $X_{6}$ & $X_{7}-\varepsilon X_{6}$ & $X_{8}$ & $%
X_{9}+\varepsilon X_{5}$ & $X_{10}$ \\
$\mathbf{X}_{6}$ & $X_{6}$ & $X_{7}$ & $X_{8}-\varepsilon X_{6}$ & $X_{9}$ &
$X_{10}$ \\
$\mathbf{X}_{7}$ & $X_{6}$ & $X_{7}$ & $X_{8}-\varepsilon X_{7}$ & $%
X_{9}-\varepsilon X_{7}$ & $X_{10}$ \\
$\mathbf{X}_{8}$ & $e^{\varepsilon }X_{6}$ & $e^{\varepsilon }X_{7}$ & $%
X_{8} $ & $X_{9}$ & $e^{\varepsilon }X_{10}$ \\
$\mathbf{X}_{9}$ & $X_{6}$ & $e^{\varepsilon }X_{7}$ & $X_{8}$ & $X_{9}$ & $%
e^{-\varepsilon }X_{10}$ \\
$\mathbf{X}_{10}$ & $X_{6}$ & $X_{7}$ & $X_{8}-\varepsilon X_{10}$ & $%
X_{9}+\varepsilon X_{10}$ & $X_{10}$ \\ \hline\hline
\end{tabular}%
\label{tab2b}%
\end{table}%

\subsection{Constant gravitational field}

We continue our discussion by introducing a constant gravitational field in
non-rotating frame, that is $f_{0}=0$. Hence, the SWMHD equations take the
following form%
\begin{eqnarray}
h_{,t}+\left( hu\right) _{,x} &=&0, \\
\left( hu\right) _{,t}+\left( hu^{2}-ha^{2}\right) _{,x}+ghh_{,x} &=&0, \\
\left( hv\right) _{,t}+\left( huv-hab\right) _{,x} &=&0, \\
\left( ha\right) _{,t}+u\left( ha\right) _{,x} &=&0, \\
\left( hb\right) _{,t}+\left( h\left( ub-va\right) \right) _{,x}+v\left(
ha\right) _{,x} &=&0.
\end{eqnarray}

Hence, the Lie symmetry condition leads to the following set of constraint
equations
\begin{equation}
\xi ^{t}=\xi ^{t}\left( t\right) ,~\xi ^{x}=\xi ^{x}\left( t,x\right) ,~\eta
^{h}=2h\left( \xi _{,x}^{x}-\xi _{,t}^{t}\right) ,
\end{equation}%
\begin{equation}
\xi _{,tt}^{t}=0,~\xi _{,tt}^{x}=0,~\xi _{,xx}^{x}=0,~\xi _{,tx}^{x}=0,
\end{equation}%
\begin{equation}
\eta ^{u}=u\left( \xi _{,x}^{x}-\xi _{,t}^{t}\right) +\xi _{,t}^{x},~\eta
^{a}=a\left( \xi _{,x}^{x}-\xi _{,t}^{t}\right) ,
\end{equation}%
\begin{equation}
a\eta _{,a}^{b}-h\eta _{,h}^{b}=0,~b\eta _{,b}^{b}-h\eta _{,h}^{b}-\eta
^{b}=0,~h\eta _{,hh}^{b}+2\eta _{,h}^{b}=0,
\end{equation}%
\begin{equation}
\eta ^{v}=\eta ^{v}\left( v\right) ,~b\eta _{,v}^{v}-h\eta _{,h}^{b}-\eta
^{b}=0.
\end{equation}

\subsubsection{Lie symmetries}

The solution of the Lie symmetry conditions for the SWMHD hyperbolic
equations within a constant gravitational field leads to the derivation of
eight Lie symmetries with form the $L^{8}$ Lie algebra.

The elements of the\ Lie algebra $L^{8}$ are
\begin{equation*}
X_{1},~X_{2},~X_{3},~X_{5},~X_{6},~X_{8},~X_{10},~Y=X^{9}-2X^{4},
\end{equation*}%
where the Galileon symmetries are preserved. Moreover, for vector field $Y$
we derive the one-parameter point transformation%
\begin{equation*}
Y:\bar{t}=e^{\varepsilon _{Y}}t,~\bar{u}=e^{-\varepsilon _{Y}}u,~\bar{a}%
=e^{-\varepsilon _{Y}}a,~\bar{h}=e^{-2\varepsilon _{4}}h.
\end{equation*}

In Tables \ref{tab3} and \ref{tab4} we present the commutators and the
adjoint representation of the Lie symmetries. \ Thus, the admitted eight Lie
symmetries form the $A_{2,1}\rtimes A_{6,22},\,\ $Lie algebra in the
Morozov-Mubarakzyanov-Patera classification scheme \cite{m1,m2,m3,m4,m5}.

\begin{table}[tbp] \centering%
\caption{Commutators of the eight admitted Lie symmetries for the SWMHD
hyperbolic system in a constant gravitational field.}%
\begin{tabular}{ccccccccc}
\hline\hline
$\left[ \mathbf{X}_{i},\mathbf{X}_{j}~\right] $ & $\mathbf{X}_{1}$ & $%
\mathbf{X}_{2}$ & $\mathbf{X}_{3}$ & $\mathbf{X}_{5}$ & $\mathbf{X}_{6}$ & $%
\mathbf{X}_{8}$ & $\mathbf{X}_{10}$ & $\mathbf{Y}$ \\
$\mathbf{X}_{1}$ & $0$ & $0$ & $X_{1}$ & $X_{2}$ & $0$ & $0$ & $0$ & $X_{1}$
\\
$\mathbf{X}_{2}$ & $0$ & $0$ & $X_{2}$ & $0$ & $0$ & $0$ & $0$ & $0$ \\
$\mathbf{X}_{3}$ & $-X_{1}$ & $-X_{2}$ & $0$ & $0$ & $0$ & $0$ & \thinspace $%
0$ & $0$ \\
$\mathbf{X}_{5}$ & $-X_{2}$ & $0$ & $0$ & $0$ & $0$ & $0$ & $0$ & $-X_{5}$
\\
$\mathbf{X}_{6}$ & $0$ & $0$ & $0$ & $0$ & $0$ & $X^{6}$ & $0$ & $0$ \\
$\mathbf{X}_{8}$ & $0$ & $0$ & $0$ & $0$ & $-X^{6}$ & $0$ & $-X_{10}$ & $0$
\\
$\mathbf{X}_{10}$ & $0$ & $0$ & $0$ & $0$ & $0$ & $X_{10}$ & $0$ & $-3X_{10}$
\\
$\mathbf{Y}$ & $-X_{1}$ & $0$ & $0$ & $X_{5}$ & $0$ & $0$ & $3X_{10}$ & $0$
\\ \hline\hline
\end{tabular}%
\label{tab3}%
\end{table}%

\begin{table}[tbp] \centering%
\caption{Adjoint representation of the eight admitted Lie symmetries for the
SWMHD hyperbolic system in a constant gravitational field.}%
\begin{tabular}{ccccccccc}
\hline\hline
$Ad\left( e^{\left( \varepsilon \mathbf{X}_{i}\right) }\right) \mathbf{X}%
_{j} $ & $\mathbf{X}_{1}$ & $\mathbf{X}_{2}$ & $\mathbf{X}_{3}$ & $\mathbf{X}%
_{5}$ & $\mathbf{X}_{6}$ & $\mathbf{X}_{8}$ & $\mathbf{X}_{10}$ & $\mathbf{Y}
$ \\
$\mathbf{X}_{1}$ & $X_{1}$ & $X_{2}$ & $X_{3}-\varepsilon X_{1}$ & $%
X_{5}-2X_{2}$ & $X_{6}$ & $X_{8}$ & $X_{10}$ & $Y-\varepsilon X_{1}$ \\
$\mathbf{X}_{2}$ & $X_{1}$ & $X_{2}$ & $X_{3}-\varepsilon X_{2}$ & $X_{5}$ &
$X_{6}$ & $X_{8}$ & $X_{10}$ & $Y$ \\
$\mathbf{X}_{3}$ & $e^{\varepsilon }X_{1}$ & $e^{\varepsilon }X_{2}$ & $%
X_{3} $ & $X_{5}$ & $X_{6}$ & $X_{8}$ & $X_{10}$ & $Y$ \\
$\mathbf{X}_{5}$ & $X_{1}+\varepsilon X_{2}$ & $X_{2}$ & $X_{3}$ & $X_{5}$ &
$X_{6}$ & $X_{8}$ & $X_{10}$ & $Y+\varepsilon X_{5}$ \\
$\mathbf{X}_{6}$ & $X_{1}$ & $X_{2}$ & $X_{3}$ & $X_{5}$ & $X_{6}$ & $%
X_{8}-\varepsilon X_{6}$ & $X_{10}$ & $Y$ \\
$\mathbf{X}_{8}$ & $X_{1}$ & $X_{2}$ & $X_{3}$ & $X_{5}$ & $e^{\varepsilon
}X_{6}$ & $X_{8}$ & $e^{\varepsilon }X_{10}$ & $Y$ \\
$\mathbf{X}_{10}$ & $X_{1}$ & $X_{2}$ & $X_{3}$ & $X_{5}$ & $X_{6}$ & $%
X_{8}-\varepsilon X_{10}$ & $X_{10}$ & $Y+3\varepsilon X_{8}$ \\
$\mathbf{Y}$ & $e^{\varepsilon }X_{1}$ & $X_{2}$ & $X_{3}$ & $%
e^{-\varepsilon }X_{5}$ & $X_{6}$ & $X_{8}$ & $e^{-3\varepsilon }X_{10}$ & $%
Y $ \\ \hline\hline
\end{tabular}%
\label{tab4}%
\end{table}%

\subsection{Coriolis force without gravitational field}

In the absence of the gravitational field, but in the presence of a constant
rotational frame the one-dimensional SWMHD equations (\ref{ww.01}), (\ref%
{ww.02}), (\ref{ww.03}), (\ref{ww.04}) and (\ref{ww.05}) read
\begin{eqnarray}
h_{,t}+\left( hu\right) _{,x} &=&0, \\
\left( hu\right) _{,t}+\left( hu^{2}-ha^{2}\right) _{,x}+f_{0}hv &=&0, \\
\left( hv\right) _{,t}+\left( huv-hab\right) _{,x}-f_{0}hu &=&0, \\
\left( ha\right) _{,t}+u\left( ha\right) _{,x} &=&0, \\
\left( hb\right) _{,t}+\left( h\left( ub-va\right) \right) _{,x}+v\left(
ha\right) _{,x} &=&0,
\end{eqnarray}

For this hyperbolic system the Lie symmetry condition provides the
constraint equations
\begin{equation}
\xi ^{t}=\xi ^{t}\left( t\right) ,~\xi ^{x}=\xi ^{x}\left( t,x\right) ,~\eta
^{h}=\eta \left( h\right) ~,~\eta ^{v}=\eta ^{v}\left( v\right) ~,~\eta
^{b}=\eta ^{b}\left( a,b,h\right) ,
\end{equation}%
\begin{equation}
\eta ^{a}=a\left( \xi _{,x}^{x}-\xi _{,t}^{t}\right) ,~\eta ^{h}=2H\left(
\xi _{,x}^{x}-\xi _{,t}^{t}\right) ,~\eta ^{u}=u\left( \xi _{,x}^{x}-\xi
_{,t}^{t}\right) +\xi _{,t}^{x},
\end{equation}%
\begin{equation}
\xi _{,tt}^{t}=0,~\xi _{,tt}^{x}=0,~\xi _{,xx}^{x}=0,~\xi _{,tx}^{x}=0,
\end{equation}%
\begin{equation}
a\eta ^{b}-h\eta _{,h}^{b}=0,~b\eta _{,b}^{b}-h\eta _{,h}^{b}-\eta ^{b}=0,
\end{equation}%
\begin{equation}
h\eta _{,hh}^{b}+2\eta _{,h}^{b}=0,~b\eta _{,v}^{v}-\eta _{,h}^{b}-\eta
^{b}=0.
\end{equation}

\subsubsection{Lie symmetries}

The solution of the Lie symmetry conditions leads to the seven dimensional
Lie algebra $L^{7}$ with elements the vector fields%
\begin{equation*}
X_{1},~X_{2},~X_{4},~X_{10},~Z_{1}=X_{3}+X_{8}-X_{9},
\end{equation*}%
\begin{equation*}
Z_{2}=\sin \left( f_{0}t\right) \partial _{x}+f_{0}\left( \cos \left(
f_{0}t\right) \partial _{u}+\sin \left( f_{0}t\right) \partial _{v}\right) ,
\end{equation*}

\begin{equation*}
Z_{3}=\cos \left( f_{0}t\right) \partial _{x}+f_{0}\left( -\sin \left(
f_{0}t\right) \partial _{u}+\cos \left( f_{0}t\right) \partial _{v}\right) .
\end{equation*}%
We remark that the Galileon symmetry is lost when the Coriolis term appears
in the system.

Thus, for the generator vector fields $Z_{1},~Z_{2}$\ and $Z_{3}$\ \ we
calculate the one-parameter point transformation%
\begin{eqnarray*}
Z_{1} &:&\bar{x}=e^{\varepsilon _{1}^{Z}}x,~\bar{u}=e^{\varepsilon
_{1}^{Z}}u,~\bar{v}=e^{\varepsilon _{1}^{Z}}v,~\bar{a}=e^{\varepsilon
_{1}^{Z}}a,~\bar{b}=e^{\varepsilon _{1}^{Z}}b, \\
Z_{2} &:&\bar{x}=x+\varepsilon _{2}^{Z}\sin \left( f_{0}t\right) ,~\bar{u}%
=u+\varepsilon _{2}^{Z}f_{0}\cos \left( f_{0}t\right) ,~\bar{v}%
=v+\varepsilon _{2}^{Z}f_{0}\sin \left( f_{0}t\right) , \\
Z_{3} &:&\bar{x}=x+\varepsilon _{3}^{Z}\cos \left( f_{0}t\right) ,~\bar{u}%
=u-\varepsilon _{3}^{Z}f_{0}\sin \left( f_{0}t\right) ,~\bar{v}%
=v+\varepsilon _{3}^{Z}f_{0}\cos \left( f_{0}t\right) ,
\end{eqnarray*}

Hence, the commutators and the adjoint representation for the Lie algebra $%
L^{7}$ are presented respectively in Tables \ref{tab5}, \ref{tab6} and \ref%
{tab6b}. Hence, in the Morozov-Mubarakzyanov-Patera classification scheme
\cite{m1,m2,m3,m4,m5}$~$it follows that $L^{7}$ is the Lie algebra $%
A_{3,5}\rtimes\left\{ A_{2,1}\rtimes A_{2,1}\right\} $.

\begin{table}[tbp] \centering%
\caption{Commutators of the seven admitted Lie symmetries for the SWMHD
hyperbolic system in a rotating frame.}%
\begin{tabular}{cccccccc}
\hline\hline
$\left[ \mathbf{X}_{i},\mathbf{X}_{j}~\right] $ & $\mathbf{X}_{1}$ & $%
\mathbf{X}_{2}$ & $\mathbf{X}_{4}$ & $\mathbf{X}_{10}$ & $\mathbf{Z}_{1}$ & $%
\mathbf{Z}_{2}$ & $\mathbf{Z}_{3}$ \\
$\mathbf{X}_{1}$ & $0$ & $0$ & $0$ & $0$ & $0$ & $f_{0}Z_{3}$ & $-f_{0}Z_{2}$
\\
$\mathbf{X}_{2}$ & $0$ & $0$ & $0$ & $0$ & $X_{2}$ & $0$ & $0$ \\
$\mathbf{X}_{4}$ & $0$ & $0$ & $0$ & \thinspace $-X_{10}$ & $0$ & $0$ & $0$
\\
$\mathbf{X}_{10}$ & $0$ & $0$ & $X_{10}$ & $0$ & $2X_{10}$ & $0$ & $0$ \\
$\mathbf{Z}_{1}$ & $0$ & $-X_{2}$ & $0$ & $-2X_{10}$ & $0$ & $-Z_{2}$ & $%
-Z_{3}$ \\
$\mathbf{Z}_{2}$ & $-f_{0}Z_{3}$ & $0$ & $0$ & $0$ & $Z_{2}$ & $0$ & $0$ \\
$\mathbf{Z}_{3}$ & $f_{0}Z_{2}$ & $0$ & $0$ & $0$ & $Z_{3}$ & $0$ & $0$ \\
\hline\hline
\end{tabular}%
\label{tab5}%
\end{table}%

\begin{table}[tbp] \centering%
\caption{Adjoint representation of the seven admitted Lie symmetries for the
SWMHD hyperbolic system in a rotating frame (1/2).}%
\begin{tabular}{ccccc}
\hline\hline
$Ad\left( e^{\left( \varepsilon \mathbf{X}_{i}\right) }\right) \mathbf{X}%
_{j} $ & $\mathbf{X}_{1}$ & $\mathbf{X}_{2}$ & $\mathbf{X}_{4}$ & $\mathbf{X}%
_{10} $ \\
$\mathbf{X}_{1}$ & $X_{1}$ & $X_{2}$ & $X_{3}$ & $X_{10}$ \\
$\mathbf{X}_{2}$ & $X_{1}$ & $X_{2}$ & $X_{3}$ & $X_{10}$ \\
$\mathbf{X}_{4}$ & $X_{1}$ & $X_{2}$ & $X_{3}$ & $e^{\varepsilon }X_{10}$ \\
$\mathbf{X}_{10}$ & $X_{1}$ & $X_{2}$ & $X_{3}-\varepsilon X_{10}$ & $X_{10}$
\\
$\mathbf{Z}_{1}$ & $X_{1}$ & $e^{\varepsilon }X_{2}$ & $X_{3}$ & $%
e^{2\varepsilon }X_{10}$ \\
$\mathbf{Z}_{2}$ & $X_{1}+\varepsilon f_{0}Z_{3}$ & $X_{2}$ & $X_{3}$ & $%
X_{10}$ \\
$\mathbf{Z}_{3}$ & $X_{1}-\varepsilon f_{0}Z_{2}$ & $X_{2}$ & $X_{3}$ & $%
X_{10}$ \\ \hline\hline
\end{tabular}%
\label{tab6}%
\end{table}%

\begin{table}[tbp] \centering%
\caption{Adjoint representation of the seven admitted Lie symmetries for the
SWMHD hyperbolic system in a rotating frame (2/2).}%
\begin{tabular}{cccc}
\hline\hline
$Ad\left( e^{\left( \varepsilon \mathbf{X}_{i}\right) }\right) \mathbf{X}%
_{j} $ & $\mathbf{Z}_{1}$ & $\mathbf{Z}_{2}$ & $\mathbf{Z}_{3}$ \\
$\mathbf{X}_{1}$ & $Z_{1}$ & $\cos \left( f_{0}\varepsilon \right)
Z_{2}-\sin \left( f_{0}\varepsilon \right) Z_{3}$ & $\sin \left(
f_{0}\varepsilon \right) Z_{2}+\sin \left( f_{0}\varepsilon \right) Z_{3}$
\\
$\mathbf{X}_{2}$ & $Z_{1}-\varepsilon X_{2}$ & $Z_{2}$ & $Z_{3}$ \\
$\mathbf{X}_{4}$ & $Z_{1}$ & $Z_{2}$ & $Z_{3}$ \\
$\mathbf{X}_{10}$ & $Z_{1}-2\varepsilon X_{10}$ & $Z_{2}$ & $Z_{3}$ \\
$\mathbf{Z}_{1}$ & $Z_{1}$ & $e^{\varepsilon }Z_{2}$ & $e^{\varepsilon
}Z_{3} $ \\
$\mathbf{Z}_{2}$ & $Z_{1}-\varepsilon Z_{2}$ & $Z_{2}$ & $Z_{3}$ \\
$\mathbf{Z}_{3}$ & $Z_{1}-\varepsilon Z_{3}$ & $Z_{2}$ & $Z_{3}$ \\
\hline\hline
\end{tabular}%
\label{tab6b}%
\end{table}%

\subsection{Constant gravitational field and Coriolis force}

For the general case of the one-dimensional hyperbolic system (\ref{ww.01}),
(\ref{ww.02}), (\ref{ww.03}), (\ref{ww.04}) and (\ref{ww.05}) with $%
gf_{0}\neq 0$, and $g,$~$f_{0}$ be constants, the Lie symmetry conditions
are expressed as follows%
\begin{equation*}
\xi ^{t}=\xi _{0}^{t}~,~\xi ^{x}=\xi ^{x}\left( t\right) ~,~\eta ^{b}=\eta
^{b}\left( a,h\right) ,
\end{equation*}%
\begin{equation*}
\xi _{,ttt}^{x}+f_{0}^{2}\xi _{,t}^{x}=0,~\xi _{,xx}^{x}=0,~\xi
_{,tx}^{x}=0,~a\eta _{,a}^{b}+\eta ^{b}-b\xi _{,x}^{x}=0,
\end{equation*}%
\begin{equation*}
\eta _{,h}^{b}-\xi _{,x}^{x}~,~h\eta _{,h}^{b}+\eta ^{b}-b\xi
_{,h}^{b}=0,~\eta ^{a}=a\xi _{,x}^{x}~,~\eta ^{h}=2h\xi _{,x}^{x},
\end{equation*}%
\begin{equation*}
\eta ^{u}=\xi _{,x}^{x}u+\xi _{,t}^{x},~\eta ^{v}=\xi _{,x}^{x}v-\frac{1}{%
f_{0}}\xi _{,tt}^{x}.
\end{equation*}

\subsubsection{Lie symmetries}

The Lie symmetries for the hyperbolic system (\ref{ww.01}), (\ref{ww.02}), (%
\ref{ww.03}), (\ref{ww.04}) and (\ref{ww.05}) with $g,~f_{0}$ nonzero
constants are
\begin{equation*}
X_{1},~X_{2},~X_{10},~Z_{1},~Z_{2},~Z_{3}.
\end{equation*}%
which form the sixth-dimensional Lie algebra $G^{6}$. \ We remark that the
Galileon symmetry is lost.

We observe that $L^{6}\subseteq L^{7}$ thus, the commutators are those
presented in Table \ref{tab5} and the adjoint representation is that given
in Table \ref{tab6}. Therefore, $L^{6}$ is the $A_{3,5}\rtimes A_{3,3}$ Lie
algebra in the Morozov-Mubarakzyanov-Patera classification scheme~\cite%
{m1,m2,m3,m4,m5}.

\section{One-dimensional optimal system for the SWMHD\ system}

\label{sec5}

We proceed with deriving the one-dimensional optimal system for the
six-dimensional Lie algebra admitted by the SWMHD system under a constant
gravitational field and a nonzero Coriolis term.

The first step in deriving the one-dimensional optimal system is to find the
invariants $\phi \left( \mathbf{a,z}\right)$ of the adjoint representation.
Based on the commutators of the Lie symmetries presented in Table \ref{tab5}%
, we define the following linear system of homogeneous partial differential
equations

\begin{eqnarray}
~z_{1}\phi _{,a_{2}} &=&0, \\
2z_{1}\phi _{,a_{10}} &=&0, \\
z_{2}\phi _{,z_{3}}-z_{3}\phi _{,z_{2}} &=&0, \\
f_{0}a_{1}\phi _{,z_{3}}-z_{1}\phi _{,z_{2}} &=&0, \\
f_{0}a_{1}\phi _{,z_{2}}+z_{1}\phi _{,z_{3}} &=&0, \\
a_{2}\phi _{,z_{2}}+2a_{10}\phi _{,a_{10}}+z_{2}\phi _{,z_{2}}+z_{3}\phi
_{,z_{3}} &=&0.
\end{eqnarray}%
Parameters $a_{1},a_{2},a_{10},z_{1},z_{2}$ and $z_{3}$ \textbf{are the
coefficient constants for the generic vector field}
\begin{equation}
\mathbf{X}%
=a_{1}X_{1}+a_{2}X_{2}+a_{10}X_{10}+z_{1}Z_{1}+z_{2}Z_{2}+z_{3}Z_{3}\text{.}
\end{equation}

The solution of the latter system is $\phi =\phi \left( a_{1},z_{1}\right) $%
, from where we infer that the invariants of the Adjoint representation are
the $a_{1}$ and $z_{1}$. This means that under the Action of the adjoint
representation the vector filed $X$ can be expressed as%
\begin{equation}
\mathbf{X}_{\left( I\right) }=a_{1}X_{1}+z_{1}Z_{1}\text{.}
\end{equation}%
On the hand, for $z_{1}=0$ and $a_{1}\neq 0$, we find $\phi =\phi \left(
a_{1},a_{10}\right) $ from where it follows that the generic vector field
can be expressed as%
\begin{equation}
\mathbf{X}_{\left( II\right) }=a_{1}X_{1}+a_{2}X_{2}+a_{10}X_{10}\text{.}
\end{equation}%
Furthermore, in the case $z_{1}\neq 0$ and $a_{1}=0$, we calculate $\phi
=\phi \left( z_{1}\right) $, then it follows%
\begin{equation}
\mathbf{X}_{\left( III\right) }=z_{1}Z_{1}\text{.}
\end{equation}%
Moreover, when $z_{1}=0$ and $a_{1}=0$, it follows $\phi =\phi \left(
a_{2},a_{10},z_{2},z_{3}\right) $, thus, the invariants are the $%
a_{2},~a_{10},~z_{2}$ and $z_{3}$ and the equivalent vector field reads%
\begin{equation}
\mathbf{X}_{\left( IV\right) }=a_{2}X_{2}+a_{10}X_{10}+z_{2}Z_{2}+z_{3}Z_{3}.
\end{equation}

From the above we conclude that the one-dimensional optimal system for the
Lie algebra $G^{6}$ is consisted by the following one-dimensional Lie
algebras%
\begin{eqnarray*}
&&\left\{ X_{1}\right\} ,~\left\{ X_{2}\right\} ,~\left\{ X_{3}\right\}
,~\left\{ X_{10}\right\} ,~\left\{ Z_{1}\right\} ,~\left\{ Z_{2}\right\}
,~\left\{ Z_{3}\right\} , \\
&&\left\{ a_{1}X_{1}+z_{1}Z_{1}\right\} ,\ \left\{
a_{1}X_{1}+a_{2}X_{2}\right\} ,~\left\{ a_{1}X_{1}+a_{10}X_{10}\right\} ,~ \\
&&\left\{ a_{1}X_{1}+a_{2}X_{2}+a_{10}X_{10}\right\} ,~\left\{
a_{2}X_{2}+a_{10}X_{10}\right\} , \\
&&\left\{ a_{2}X_{2}+z_{2}Z_{2}\right\} ,~\left\{
a_{2}X_{2}+z_{3}Z_{3}\right\} ,~\left\{ a_{10}X_{10}+z_{2}Z_{2}\right\} , \\
&&\left\{ a_{10}X_{10}+z_{3}Z_{3}\right\} ,~\left\{
z_{2}Z_{2}+z_{3}Z_{3}\right\} ,~\left\{
a_{2}X_{2}+a_{10}X_{10}+z_{2}Z_{2}\right\} , \\
&&\left\{ a_{2}X_{2}+a_{10}X_{10}+z_{3}Z_{3}\right\} ,~\left\{
a_{10}X_{10}+z_{2}Z_{2}+z_{3}Z_{3}\right\} , \\
&&\left\{ a_{2}X_{2}+a_{10}X_{10}+z_{2}Z_{2}+z_{3}Z_{3}\right\} .
\end{eqnarray*}

\section{Similarity solutions}

\label{sec6}

We continue our analysis by constructing similarity solutions derived from
the identified Lie symmetries. For each symmetry vector, we define a
similarity transformation that can reduce the hyperbolic system (\ref{ww.01}%
), (\ref{ww.02}), (\ref{ww.03}), (\ref{ww.04}), and (\ref{ww.05}) to a
system of ordinary differential equations.

In the following, we use elements of the one-dimensional optimal system to
define the similarity transformations, write the reduced system, and
determine when a closed-form solution can be derived.

\subsection{Reduction with $X_{1}$}

The Lie symmetry vector $X_{1}$ leads to the construction of the invariants $%
x,$ $h,$ $u,~v,~a,$ $b\,$. Hence, we define the similarity transformation%
\begin{equation}
h=h\left( x\right) ,~u=u\left( x\right) ,~v=v\left( x\right) ,~a=a\left(
x\right) ,~b=b\left( x\right) ,
\end{equation}%
which describes static solution.

By replacing in (\ref{ww.01}), (\ref{ww.02}), (\ref{ww.03}), (\ref{ww.04})
and (\ref{ww.05}) we end with the following system or ordinary differential
equations%
\begin{eqnarray}
\left( hu\right) _{,x} &=&0, \\
\left( hu^{2}+\frac{g}{2}h^{2}-ha^{2}\right) _{,x}+f_{0}hv &=&0, \\
\left( huv-hab\right) _{,x}-f_{0}hu &=&0, \\
u\left( ha\right) _{,x} &=&0, \\
\left( h\left( ub-va\right) \right) _{,x}+v\left( ha\right) _{,x} &=&0,
\end{eqnarray}

For $u\neq 0$, it follows that
\begin{equation}
u=\frac{u_{0}}{h},~v=-\frac{f_{0}u_{0}a_{0}}{a_{0}^{2}-u_{0}^{2}}t+v_{0},
\end{equation}%
\begin{equation}
a=\frac{a_{0}}{h},~b=-\frac{f_{0}u_{0}^{2}}{a_{0}^{2}-u_{0}^{2}}t+b_{0},
\end{equation}%
while $h\left( t\right) $ satisfies the nonlinear differential equation%
\begin{equation}
\frac{\left( gh^{3}+a_{0}^{2}-u_{0}^{2}\right) }{h}h_{,t}-f_{0}^{2}\frac{%
f_{0}u_{0}^{2}}{a_{0}^{2}-u_{0}^{2}}t+v_{0}h=0.
\end{equation}%
The latter solution exist for initial conditions in which $%
a_{0}^{2}-u_{0}^{2}\neq 0$ and $u_{0}\neq 0$.

\subsection{Reduction with $X_{2}$}

The Lie symmetry vector $X_{2}$ leads to the similarity transformation%
\begin{equation}
h=h\left( t\right) ,~u=u\left( t\right) ,~v=v\left( t\right) ,~a=a\left(
t\right) ,~b=b\left( t\right) ,
\end{equation}%
which describes stationary solution.

Thus we derive the reduced system described by the following set of ordinary
differential equations

\begin{eqnarray}
h_{,t} &=&0, \\
\left( hu\right) _{,t}+f_{0}hv &=&0, \\
\left( hv\right) _{,t}-f_{0}hu &=&0, \\
\left( ha\right) _{,t} &=&0, \\
\left( hb\right) _{,t} &=&0,
\end{eqnarray}%
from where it follows%
\begin{equation}
h=h_{0},~a=a_{0},~b=b_{0},
\end{equation}%
and%
\begin{eqnarray}
u\left( t\right) &=&u_{0}\cos \left( f_{0}t\right) -v_{0}\sin \left(
f_{0}t\right) , \\
v\left( t\right) &=&u_{0}\sin \left( f_{0}t\right) +v_{0}\cos \left(
f_{0}t\right) .
\end{eqnarray}

\subsection{Reduction with $X_{3}$}

The application of the Lie symmetry $X_{3}$ gives the similarity
transformation%
\begin{equation}
\sigma =\frac{x}{t},~h=h\left( \sigma \right) ,~u=u\left( \sigma \right)
,~v=v\left( \sigma \right) ,~a=a\left( \sigma \right) ,~b=b\left( \sigma
\right) ,
\end{equation}%
from where it follows the constant solution%
\begin{equation}
h=h_{0},~u=u_{0},~a=a_{0},~v=0\text{,~}b=0.
\end{equation}

\subsection{Reduction with $Z_{1}$}

From the Lie symmetry vector $Z_{1}$ we construct the similarity
transformation%
\begin{equation}
h=x^{2}H\left( t\right) ,~u=xU\left( t\right) ,~v=xV\left( t\right)
,~a=xA\left( t\right) ,~b=xB\left( t\right) \text{.}
\end{equation}%
Therefore, the hyperbolic system (\ref{ww.01}), (\ref{ww.02}), (\ref{ww.03}%
), (\ref{ww.04}) and (\ref{ww.05}) is reduced as follows%
\begin{eqnarray}
H_{,t}+3UH &=&0, \\
\left( HU\right) _{,t}+H\left( 4\left( U^{2}-A^{2}\right) +2gH+f_{0}V\right)
&=&0, \\
\left( HV\right) _{,x}+H\left( 4\left( UV-AB\right) -f_{0}U\right) &=&0, \\
\left( HA\right) _{,t}+3UAH &=&0, \\
\left( HB\right) _{,t}+H\left( 4UB-VA\right) &=&0.
\end{eqnarray}

Thus, by replacing $H\left( t\right) =H_{0}e^{-3\int U\left( t\right) dt}$
the latter system is simplified as%
\begin{eqnarray}
U_{,t}+U^{2}-4A+f_{0}V+2ge^{-3\int Udt} &=&0, \\
V_{,t}+UV-4AB-f_{0}U &=&0, \\
A_{,t} &=&0, \\
B_{,t}+\left( 4UB-VA\right) &=&0.
\end{eqnarray}%
The analytic solution is expressed in elliptic integrals.

\subsection{Reduction with $Z_{2}$}

The Lie symmetry vector $Z_{2}$ provides the similarity transformation%
\begin{equation}
h=h\left( t\right) ,~u=\cot \left( f_{0}t\right) f_{0}x+U\left( t\right)
,~v=f_{0}x+V\left( t\right) ,~a=a\left( t\right) ,~b=b\left( t\right) .
\end{equation}%
Hence, the SWMHD equations provides the closed-form solution%
\begin{equation}
h\left( t\right) =\frac{h_{0}}{\sin \left( f_{0}t\right) },~V\left( t\right)
=V_{0},~a=a_{0}\sin \left( f_{0}t\right) ,~b=-b_{0}\cos \left( f_{0}t\right)
+b_{1},
\end{equation}%
and%
\begin{equation}
U\left( t\right) =\frac{U_{0}}{\sin \left( f_{0}t\right) }-h_{0}\cot \left(
f_{0}t\right) .
\end{equation}

\textbf{The solution describes a shock wave with wall }$f_{0}t=n\pi ,~n\in
\mathbb{N}
$\textbf{.}

\subsection{Reduction with $Z_{3}$}

The Lie symmetry vector $Z_{3}$ provides the similarity transformation%
\begin{equation}
h=h\left( t\right) ,~u=-\tan \left( f_{0}t\right) f_{0}x+U\left( t\right)
,~v=f_{0}x+V\left( t\right) ,~a=a\left( t\right) ,~b=b\left( t\right) .
\end{equation}%
Hence, the SWMHD equations provides the closed-form solution%
\begin{equation}
h\left( t\right) =\frac{h_{0}}{\cos \left( f_{0}t\right) },~V\left( t\right)
=V_{0},~a=a_{0}\cos \left( f_{0}t\right) ,~b=-b_{0}\sin \left( f_{0}t\right)
+b_{1},
\end{equation}%
and%
\begin{equation}
U\left( t\right) =\frac{U_{0}}{\cos \left( f_{0}t\right) }-h_{0}\tan \left(
f_{0}t\right) .
\end{equation}

\textbf{The solution describes a shock wave with wall }$f_{0}t=n\frac{\pi }{2%
},~n\in
\mathbb{N}
$\textbf{. } \textbf{The similarity solutions }$Z_{2}$\textbf{\ and }$Z_{3}$%
\textbf{\ are related through the change of variables }$f_{0}t\rightarrow
f_{0}t+\frac{\pi }{4}$\textbf{. }

\subsection{Reduction with $X_{1}+a_{2}X_{2}$}

The Lie symmetry $X_{1}+a_{2}X_{2}$ provides the similarity transformation%
\begin{equation}
\xi =x-a_{2}t,~h=h\left( \xi \right) ,~u=u\left( \xi \right) ,~v=v\left( \xi
\right) ,~a=a\left( \xi \right) ,~b=b\left( \xi \right) ,
\end{equation}%
where $a_{2}$ is the speed of the travel-waves.

\begin{figure}[tbph]
\centering\includegraphics[width=1\textwidth]{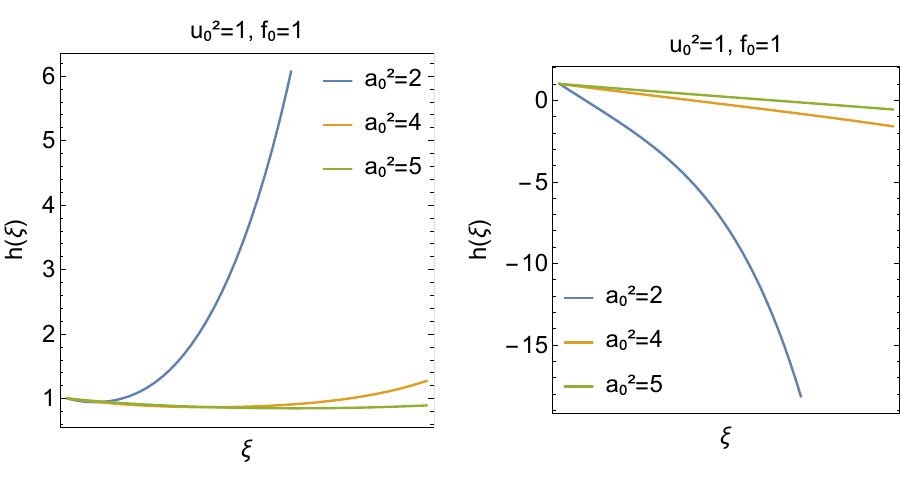}
\caption{Qualitative evolution of the functions $h\left( \protect\xi \right)
$ and $v\left( \protect\xi \right) $ as they are given by numerical
simulation of the nonlinear system (\protect\ref{dd.01}), (\protect\ref%
{dd.02}).}
\label{fig1}
\end{figure}

By replacing in the SWMHD equations we end with the system%
\begin{equation}
u=a_{2}+\frac{u_{0}}{h\left( \xi \right) },~a=\frac{a_{0}}{h\left( \xi
\right) },~b=\frac{a_{0}}{u_{0}}v\left( \xi \right) +b_{0},
\end{equation}%
\begin{eqnarray}
u_{0}\left( 1-\left( \frac{a_{0}}{u_{0}}\right) ^{2}\right) v_{,\xi
}-f_{0}\left( a_{2}h+u_{0}\right)  &=&0,  \label{dd.01} \\
\frac{\left( a_{0}^{2}-u_{0}^{2}\right) }{h^{2}}h_{,\xi }+\frac{1}{2}g\left(
h^{2}\right) _{,\xi }+f_{0}vh &=&0.  \label{dd.02}
\end{eqnarray}%
The solution is expressed in elliptic integrals. The numerical solution of
the system (\ref{dd.01}), (\ref{dd.02}) is given in Fig. \ref{fig1}.

\subsection{Reduction with $X_{2}+z_{2}Z_{2}$}

From the vector field $X_{2}+z_{2}Z_{2}$ we define the similarity
transformation%
\begin{equation}
\zeta =x+\frac{z_{2}}{f_{0}}\cos \left( f_{0}t\right) ,~h=h\left( \zeta
\right) ,~a=a\left( \zeta \right) ,~b=b\left( \zeta \right) ,
\end{equation}%
\begin{equation}
u=z_{2}\sin \left( f_{0}t\right) +U\left( \zeta \right) ,~v=-z_{2}\cos
\left( f_{0}t\right) +V\left( \zeta \right) .
\end{equation}

Therefore the hyperbolic system (\ref{ww.01}), (\ref{ww.02}), (\ref{ww.03}),
(\ref{ww.04}) and (\ref{ww.05}) provides%
\begin{equation}
a=\frac{a_{0}}{h},~U=\frac{u_{0}}{h},~b=-\frac{f_{0}a_{0}u_{0}}{%
a_{0}^{2}-u_{0}^{2}}\zeta +b_{0}~
\end{equation}%
\begin{equation}
V=-\frac{f_{0}u_{0}^{2}}{a_{0}^{2}-u_{0}^{2}}\zeta +v_{0},
\end{equation}%
and%
\begin{equation}
\frac{1}{2}\left( \frac{1}{h^{2}}\right) _{,\zeta }\left(
u_{0}^{2}-a_{0}^{2}-gh^{3}\right) -\frac{f_{0}^{2}u_{0}^{2}}{%
a_{0}^{2}-u_{0}^{2}}\zeta +f_{0}v_{0}=0.  \label{sp.01}
\end{equation}

\begin{figure}[tbph]
\centering\includegraphics[width=0.5\textwidth]{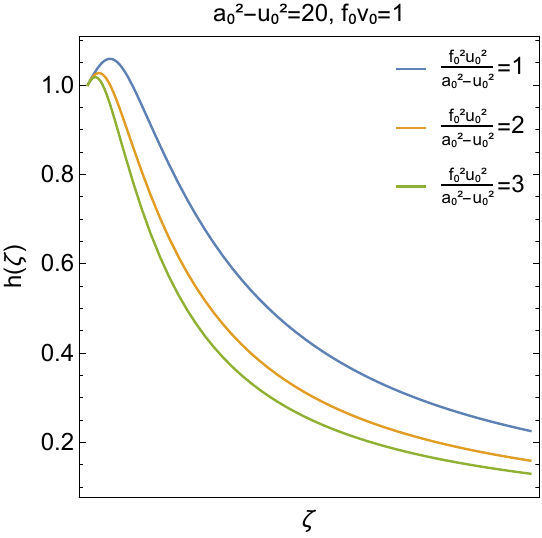}
\caption{Qualitative evolution for the function $h\left( \protect\zeta %
\right) $ as it is given by numerical simulation of the nonlinear
differential equation (\protect\ref{sp.01}). }
\label{fig2}
\end{figure}

Numerical simulations of the latter nonlinear ordinary differential
equations are presented in Fig. \ref{fig2}

\subsection{Reduction with $X_{10}+z_{2}Z_{2}$}

From the Lie symmetry vector $X_{10}+z_{2}Z_{2}$ we find the similarity
transformation%
\begin{equation}
h=h\left( t\right) ,~u=\cot \left( f_{0}t\right) f_{0}x+U\left( t\right)
,~v=f_{0}x+V\left( t\right) ,
\end{equation}%
\begin{equation}
a=a\left( t\right) ,~b=\frac{x}{z_{2}\sin \left( f_{0}t\right) h\left(
t\right) a\left( t\right) }+B\left( t\right) .
\end{equation}%
Thus, from the SWMHD equations it follows%
\begin{equation}
h\left( t\right) =\frac{h_{0}}{\sin \left( f_{0}t\right) },~V\left( t\right)
=\frac{t}{h_{0}z_{2}}+V_{0},
\end{equation}%
\begin{equation}
a\left( t\right) =a_{0}\sin \left( f_{0}t\right) ,~
\end{equation}%
\begin{equation*}
U\left( t\right) =\frac{U_{0}}{\sin \left( f_{0}t\right) }+\frac{1}{%
h_{0}z_{2}}\left( \cot \left( f_{0}t\right) \left( t+h_{0}V_{0}z_{2}\right) -%
\frac{1}{f_{0}}\right) ,
\end{equation*}%
and%
\begin{equation}
B\left( t\right) =-a_{0}\cos \left( f_{0}t\right) +\frac{t}{\left(
h_{0}z_{2}\right) ^{2}a_{0}\sin \left( f_{0}t\right) }+\frac{\left(
V_{0}+U_{0}\cos \left( f_{0}t\right) \right) }{h_{0}z_{2}a_{0}\sin \left(
f_{0}t\right) }+B_{0}.
\end{equation}

\textbf{We observe that this solution describes shoch waves with wall at }$%
f_{0}t=n\pi $\textbf{.}

\subsection{Reduction with $X_{10}+z_{3}Z_{3}$}

From the application of the Lie symmetry vector $X_{10}+z_{3}Z_{3}$ we
derive the similarity transformation%
\begin{equation}
h=h\left( t\right) ,~u=-\tan \left( f_{0}t\right) f_{0}x+U\left( t\right)
,~v=f_{0}x+V\left( t\right) ,
\end{equation}%
\begin{equation}
a=a\left( t\right) ,~b=\frac{x}{z_{3}\cos \left( f_{0}t\right) h\left(
t\right) a\left( t\right) }+B\left( t\right) .
\end{equation}%
Thus, from the SWMHD equations it follows%
\begin{equation}
h\left( t\right) =\frac{h_{0}}{\cos \left( f_{0}t\right) },~V\left( t\right)
=\frac{t}{h_{0}z_{2}}+V_{0},
\end{equation}%
\begin{equation}
a\left( t\right) =a_{0}\cos \left( f_{0}t\right) ,~
\end{equation}%
\begin{equation}
U\left( t\right) =\frac{U_{0}}{\cos \left( f_{0}t\right) }-\frac{1}{%
h_{0}z_{3}}\left( \tan \left( f_{0}t\right) \left( t+h_{0}V_{0}z_{2}\right) +%
\frac{1}{f_{0}}\right) ,
\end{equation}%
and%
\begin{equation}
B\left( t\right) =a_{0}\sin \left( f_{0}t\right) +\frac{t}{\left(
h_{0}z_{3}\right) ^{2}a_{0}\cos \left( f_{0}t\right) }+\frac{\left(
V_{0}-U_{0}\sin \left( f_{0}t\right) \right) }{h_{0}z_{2}a_{0}\cos \left(
f_{0}t\right) }+B_{0}.
\end{equation}%
We observe that this solution describe solitons. \textbf{We observe that
this solution describes shoch waves with wall at }$f_{0}t=n\pi $\textbf{.
The latter two solutions are related through the transformation }$%
f_{0}t\rightarrow f_{0}t+\frac{\pi }{4}$.

\subsection{Reduction with $X_{2}+a_{10}X_{10}+z_{2}Z_{2}$}

The generic vector field $X_{2}+a_{10}X_{10}+z_{2}Z_{2}$ leads to the
derivation of the similarity transformation%
\begin{equation}
h=h\left( t\right) ,~u=\frac{z_{2}\cos \left( f_{0}t\right) f_{0}}{%
1+z_{2}\cos \left( f_{0}t\right) }x+U\left( t\right) ,~v=\frac{z_{2}\sin
\left( f_{0}t\right) }{1+z_{2}\sin \left( f_{0}t\right) }+V\left( t\right) ,
\end{equation}%
\begin{equation}
a=a\left( t\right) ,~b=\frac{x}{h\left( t\right) a\left( t\right) \left(
1+z_{2}\cos \left( f_{0}t\right) \right) }+B\left( t\right) .
\end{equation}%
By replacing \ the later transformation in the hyperbolic SWMHD (\ref{ww.01}%
), (\ref{ww.02}), (\ref{ww.03}), (\ref{ww.04}) and (\ref{ww.05}) system we
determine the closed-form solution%
\begin{equation}
h\left( t\right) =\frac{h_{0}}{1+z_{2}\sin \left( f_{0}t\right) },~
\end{equation}%
\begin{equation*}
a\left( t\right) =a_{0}\left( 1+z_{2}\sin \left( f_{0}t\right) \right) ,
\end{equation*}%
\begin{equation}
V\left( t\right) =\frac{\sin \left( f_{0}t\right) V_{0}+\cos \left(
f_{0}t\right) U_{0}}{1+z_{2}\sin \left( f_{0}t\right) }+\frac{a_{10}z_{2}}{%
h_{0}}\frac{t}{\left( 1+z_{2}\sin \left( f_{0}t\right) \right) },
\end{equation}%
\begin{equation}
U\left( t\right) =\frac{f_{0}h_{0}\left( U_{0}\cos \left( f_{0}t\right)
+V_{0}\sin \left( f_{0}t\right) \right) +a_{10}z_{2}\sin \left(
f_{0}t\right) }{h_{0}f_{0}\left( 1+z_{2}\sin \left( f_{0}t\right) \right) },
\end{equation}%
and the magnetic field $B\left( t\right) $ is given by the first-order
linear ordinary differential equation%
\begin{equation}
B_{,t}=-\frac{h_{,t}}{h}B-\frac{a_{10}U}{ha\left( 1+z_{2}\sin \left(
f_{0}t\right) \right) }+\frac{z_{2}^{2}f_{0}\left( V\sin \left(
2f_{0}t\right) -2a\sin \left( f_{0}t\right) ^{2}\right) +z_{2}f_{0}\left(
a\sin \left( f_{0}t\right) -V\cos \left( f_{0}t\right) \right) }{2\left(
1-z_{2}\sin ^{2}\left( f_{0}t\right) \right) }.
\end{equation}%
When $\left\vert z_{2}\right\vert \leq 1$, then solitons exist.

Moreover, the vector field $X_{2}+a_{10}X_{10}+z_{3}Z_{3}$ leads to the
derivation of a similar similarity solution which can derived after a phase
transition. \

\begin{figure}[tbph]
\centering\includegraphics[width=1\textwidth]{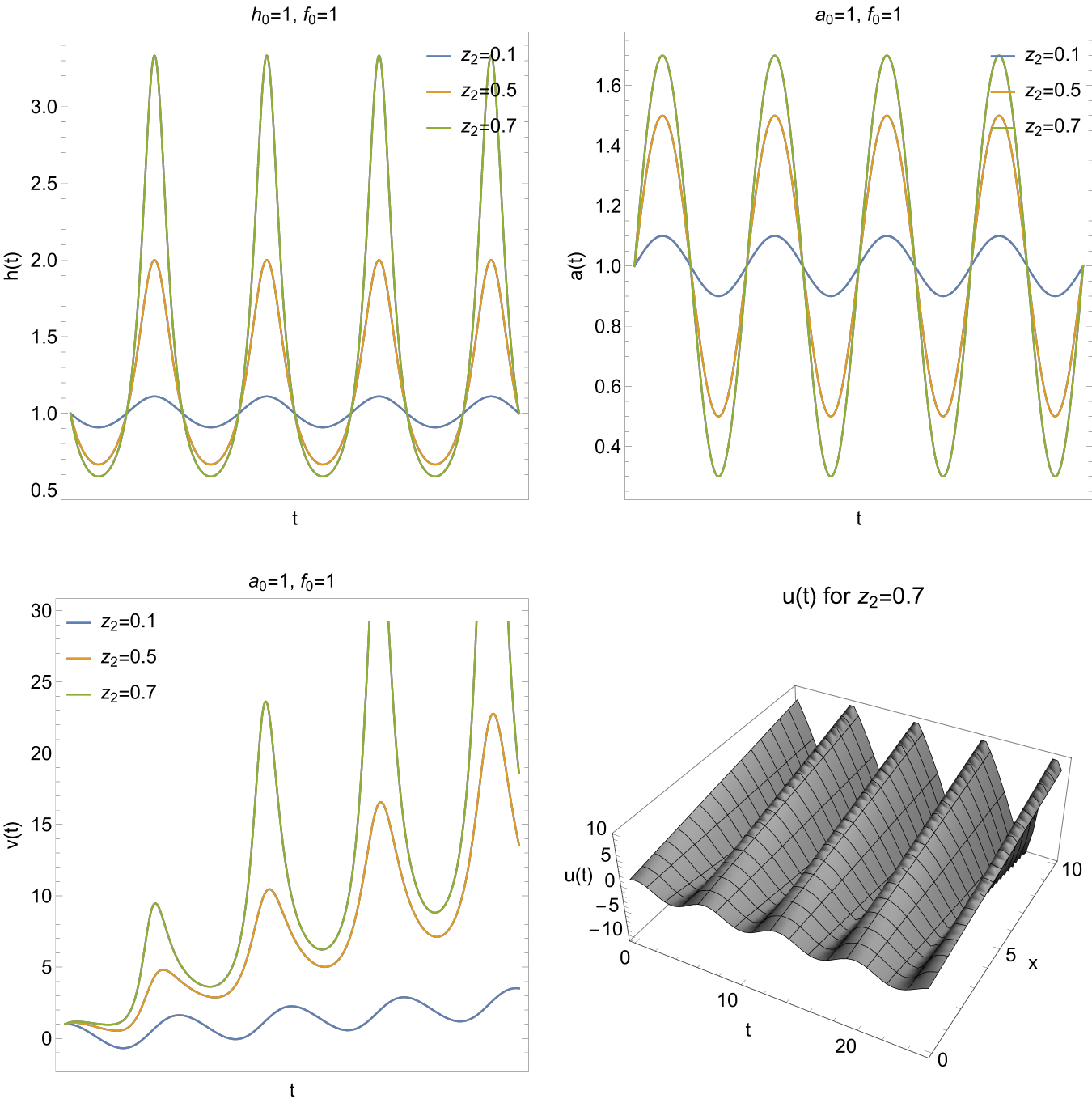}
\caption{Qualitative evolution for the similarity solution of $h\left(
t\right) ,~a\left( t\right) ,~V\left( t\right) $,~$U\left( t\right) $ as
they are given by the application of the Lie symmetry vector $%
X_{2}+a_{10}X_{10}+z_{2}Z_{2}$. }
\label{fig3}
\end{figure}

\textbf{In Fig. \ref{fig3} we present the qualitative evolution ofor the
similarity solution derived before.\ The plots are for unitary values of the
free parameters. }

\section{Conclusions}

\label{sec7}

In this work, we investigated the Lie symmetries of a hyperbolic system
consisting of five first-order nonlinear partial differential equations that
describe an SWMHD system. Hyperbolic systems are crucial for describing
various physical phenomena. Symmetry analysis is a powerful tool for the
analytical study of nonlinear differential equations, and Lie symmetries
offer a systematic approach to better understand these systems.

In the context of the SW equations, Lie symmetry analysis has been employed
to determine analytic solutions and conservation laws. Similarly, Lie theory
has played a significant role in the study of MHD systems. We considered a
one-dimensional SWMHD model with the relaxing condition $\nabla \left( h%
\mathbf{B}\right) \neq 0$, which is physically acceptable as long as the
boundary condition $\nabla \left( h\mathbf{B}\right) =0$ is met.

The introduction of the term $\nabla \left( h\mathbf{B}\right) $\ restores
Galilean symmetries in the free SWMHD system without the Coriolis term. We
found that the admitted symmetries number ten and form the Lie algebra $%
A_{3,3}\rtimes A_{2,1}\rtimes A_{5,34}^{a}$. When a constant gravitational
field is present, the number of admitted Lie symmetries reduces to eight,
and the Lie algebra is $A_{2,1}\rtimes A_{6,22}$. These results differ from
those of the one-dimensional SWMHD system where the condition $\nabla \left(
h\mathbf{B}\right) =0$\ is imposed.

Furthermore, when a Coriolis force term appears, related to the rotating
reference frame, the number of admitted Lie symmetries is further reduced to
seven or six, depending on the presence of the gravitational field. The
corresponding Lie algebras are $A_{3,5}\rtimes A_{2,1}\rtimes A_{2,1}$ and $%
A_{3,5}\rtimes A_{3,3}$, respectively. For the most general case, where the
Lie symmetries form the $A_{3,5}\rtimes A_{3,3}$ Lie algebra, we calculated
the one-dimensional optimal system and used similarity transformations to
reduce the hyperbolic system to a system of ordinary differential equations,
from which we derived the analytic solution. The one-dimensional optimal
system was obtained by determining all the invariants of the adjoint action
for the $A_{3,5}\rtimes A_{3,3}$ Lie algebra and its subalgebras.

In this study, we have used symmetries to construct invariants that are
primarily applied to reduce the system of hyperbolic equations.
Nevertheless, another important feature of symmetry analysis is that it can
be used to identify conservation laws \cite{nn1}. There are different
approaches in the literature regarding how to use symmetries to identify
conservation laws. For systems that follow from a variational principle,
Noether's theorem is mainly applied. Another approach is Ibragimov's
conservation law theorem \cite{nn2}. For other studies, we refer the reader
to \cite{nn3,nn4,nn5} and the references therein.

In future work, we plan to further investigate the algebraic properties of
this specific SWMHD model by applying the theory of higher-order symmetries
and determining conservation laws related to the symmetry vectors.
Additionally, the symmetry analysis of the SWMHD system in Lagrangian
coordinates will be of special interest.

\textbf{Data Availability Statements:} Data sharing is not applicable to
this article as no datasets were generated or analyzed during the current
study.

\textbf{Code Availability Statements:} Code sharing is not applicable to
this article as no code was used in this study.

\textbf{Acknowledgements:}AP thanks the support of Vicerrector\'{\i}a de
Investigaci\'{o}n y Desarrollo Tecnol\'{o}gico (Vridt) at Universidad Cat%
\'{o}lica del Norte through N\'{u}cleo de Investigaci\'{o}n Geometr\'{\i}a
Diferencial y Aplicaciones, Resoluci\'{o}n Vridt No - 096/2022 and Resoluci%
\'{o}n Vridt No - 098/2022. Part of this work was supported by Proyecto
Fondecyt Regular 2024, Folio 1240514, Etapa 2024. AP thanks AH and the
Woxsen University for the hospitality provided while part of this work was
performed.


\begin{thebibliography}{99}
\bibitem{nn5} S.C. Anco, On the incompleteness of Ibragimov's conservation
law theorem and its equivalence to a standard formula using symmetries and
adjoint-symmetries,Symmetry 9, 33 (201)
\end{thebibliography}
\end{document}